\documentclass[reprint,showpacs,showkeys,preprintnumbers,amsmath,amssymb,aps,prl]{revtex4-1}

\usepackage{graphicx}% Include figure files
\usepackage{amsmath}
\usepackage{amssymb}
\begin{document}

\preprint{Y. Q. Zhang \textit{et al.}, Redefinition of spin Hall magnetoresistance}

\title{Redefinition of spin Hall magnetoresistance}

\author{Yan-Qing Zhang$^{1}$}
%\affiliation{School of Physics Science and Engineering, Tongji University, Shanghai 200092, China.}

\author{Hua-Rui Fu$^{2}$}

\author{Niu-Yi Sun$^{1}$}
%\affiliation{School of Physics Science and Engineering, Tongji University, Shanghai 200092, China.}

\author{Wen-Ru Che$^{1}$}
%\affiliation{School of Physics Science and Engineering, Tongji University, Shanghai 200092, China.}

\author{Ding Ding$^{3}$}
%\affiliation{Institute of Materials, Shanghai University, Shanghai 200072, China.}

\author{Juan Qin$^{3}$}
\email{juan\_qin@staff.shu.edu.cn}
\author{Cai-Yin You$^{2}$}
\email{caiyinyou@xaut.edu.cn}
\author{Zhen-Gang Zhu$^{4}$}
\email{zgzhu@ucas.ac.cn}
%\affiliation{School of Electronic, Electrical and Communication Engineering, University of Chinese Academy of Sciences, Beijing 100049, China.}

\author{Rong Shan$^{1}$}
%\email{shan.rong@hotmail.com}
%\affiliation{School of Physics Science and Engineering, Tongji University, Shanghai 200092, China.}

\affiliation{
$^{1}$ Shanghai Key Laboratory of Special Artificial Microstructure Materials and Technology$\&$School of Physics Science and Engineering, Tongji University, Shanghai 200092, China.\\
%$^{2}$ Institute of Materials, Shanghai University, Shanghai 200072, China\\
$^{2}$ School of Materials Science and Engineering, Xi'an University of Technology, Xi'an 710048, China\\
$^{3}$ School of Materials Science and Engineering, Shanghai University, Shanghai 200072, China\\
$^{4}$ School of Electronic, Electrical and Communication Engineering, University of Chinese Academy of Sciences, Beijing 100049, China.\\
%$^{5}$ Theoretical Condensed Matter Physics and Computational Materials Physics Laboratory, College of Physical Sciences, University of Chinese Academy of Sciences, Beijing 100049, China.
}

\date{\today}

\begin{abstract}
Using a multi-conduction-channel model, we redefined the micromechanism of spin Hall magnetoresistance (SMR). Four conduction channels are created by spin accumulation of nonpolarized electron flow at top, bottom, left and right interfaces of the film sample, which corresponds to different resistance states of polarized electron flow with various spin directions relative to the applied magnetic field ($\mathbf{H}$), and brings about the SMR effect finally. The magnetic insulator layer, such as yttrium iron garnet (YIG), is not a requisite for the observation of SMR. Instead, the SMR effect is perfectly realized, with an order of magnitude increase, in the sample with a discontinuous layer of isolated-Co$_2$FeAl (0.3 nm) grains covered by 2.5-nm-thick Pt layer on MgO substrate. The model intuitively gives the typical relationship of SMR effect, i.e. $\rho_{\parallel}\approx\rho_{\bot}>\rho_{T}$, where $\rho_{\bot}$, $\rho_{\parallel}$ and $\rho_{T}$ are longitudinal reisitivities with applied magnetic field ($\mathbf{H}$) direction perpendicular to the current direction out of plane (as Z direction), parallel with and perpendicular to it in plane (as X and Y direction), respectively. Our research reveals that the scattering between polarized and nonpolarized conduction electrons is the origin of SMR, and the intrinsic SMR is not constant when $\mathbf{H}$ direction rotates in XZ plane, which is distinct from that in the reported SMR mechanism. 

\end{abstract}


%\keywords{proximity effect, spin orbit coupling, anomalous Hall effect, anomalous Nernst effect, Hall angle}

\maketitle
%\section{Introduction} %%%%%%%%%%%%%%%%%%%%%%%%%%%%%%%%%%%%%%%%%
Since the magnetic transport property of Pt grown on yttrium iron garnet (YIG) was reported two years ago~\cite{Huang2012}, the drastic controversy on the understanding of its origin continues today. On the beginning, the strong ferromagnetic characteristics in Pt films on YIG are considered as a consequence of static magnetic proximity effect (MPE)~\cite{Huang2012,Bergmann1978,Wilhelm2000,Lu2013}. Soon, Nakayama \textit{et al}. excluded the contribution of static magnetic proximity effect through an inserted 6-nm-thick Cu layer between Pt and YIG, where the magnetic transport property of YIG/Cu/Pt film persists although it turns much weak~\cite{Nakayama2013}. They pointed out that the magnetic transport property of Pt/YIG is actually induced by spin Hall effect and meantime suggested a new magnetoresistance (MR) phenomenon, i.e. spin Hall magnetoresistance (SMR). Next, Lu Y. M. \textit{et al}. insist on the MPE by a series of experiments on YIG/Pt and Pt/Permalloy/Pt fims~\cite{Lu2013PRB}. They found $\rho_{\parallel}\approx\rho_{\bot}>\rho_{T}$, where $\rho_{\bot}$, $\rho_{\parallel}$ and $\rho_{T}$ are longitudinal reisitivities with applied magnetic field ($\mathbf{H}$) direction perpendicular to the current direction out of plane (as Z direction), parallel with and perpendicular to it in plane (as X and Y direction), respectively. The behavior is distinctly different from all other known MR effects including the SMR, and termed the new MR as hybrid MR. Immediately, the proposal of hybrid MR caused a more intense argument~\cite{Miao2014,Lin2014,Althammer2013,Chen2013,Huang2014,Li2014,Kobs2014}. To understand the nature of MR in YIG/Pt films, Lin T. \textit{et al}.~\cite{Miao2014} and Miao B. F. \textit{et al}.~\cite{Lin2014} suggested two compromised solutions. The former appraises that the SMR plays a main role at high temperature and MPE gradually manifests itself with decreasing temperature. The latter believes that the SMR dominates at low magnetic field, while MPE at high field. In review all these arguments, we found all viewpoints, to some degree, has its reasonable and unreasonable aspects. However, \textit{Numquam ponenda est pluralitas sine necessitate}, as Occam had said hundreds years ago, there might be a model to comprehend all experimental results simply. Under this consideration, we therefore redefined the SMR in this work.
% Figure 1 %%%%%%%%%%%%%%%%%%%%%%%%%%%%%%%%%%%%%%%%%%%%%%%%%%%%%%%%%
\begin{figure}
\centering
   \includegraphics[width=8.5cm]{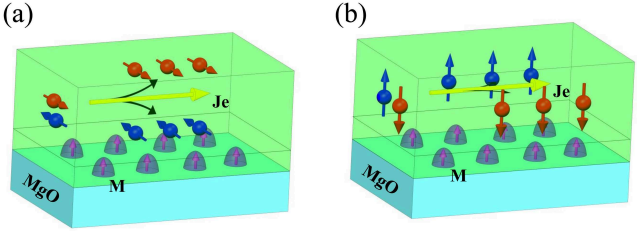}
   \caption{(Color online)  Spin Hall effect in a nonmagnetic metal layer grown on a discontinuous ferromagnetic layer. Spin accumulation occurs at: (a) top and bottom interfaces; (b) left and right interfaces.}
\label{Fig 1}
\end{figure}
%%%%%%%%%%%%%%%%%%%%%%%%%%%%%%%%%%%%%%%%%%%%%%%%%%%%%%%%%%%%%%%%%%%%
% Figure 2 %%%%%%%%%%%%%%%%%%%%%%%%%%%%%%%%%%%%%%%%%%%%%%%%%%%%%%%%%
\begin{figure*}
\centering
%\begin{minipage}{0.6\textwidth}
   \includegraphics[width=10.5cm]{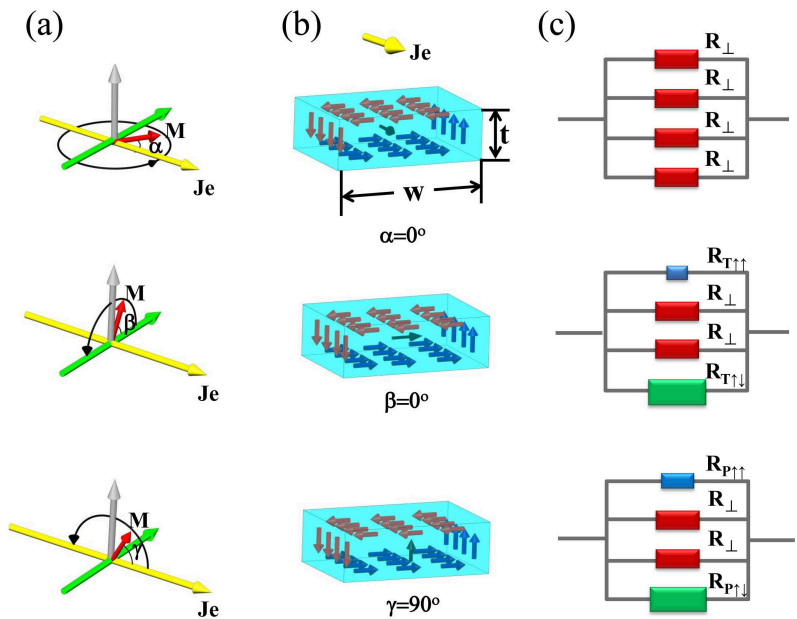}
   \caption{(Color online)  Micromechanism of spin Hall magnetoresistance. (a) Coordinate system of the electron flow (yellow arrow), transverse direction (green arrow) in plane, perpendicular direction (gray arrow), and schematic diagram notations of rotation angles of magnetic field in different planes. (b) Four spin channels at top, bottom, left and right interfaces, generated by spin Hall effect of nonpolarized electron flow. At the center of every image, black arrow represents spin direction of polarized electrons along the applied magnetic field. (c) Equivalent circuit of spin hall magnetoresistance when $\alpha$=$0~^\circ$, $\beta$=$0~^\circ$ and $\gamma$=$90~^\circ$, i.e. $\rho_{\parallel}$, $\rho_{T}$ and $\rho_{\bot}$. $R_\bot$, $R_{T\uparrow\uparrow}$ ($R_{P\uparrow\uparrow}$) and $R_{T\uparrow\downarrow}$ ($R_{P\uparrow\downarrow}$) stand for resistivity states of polarized electron flow passing through different channels, respectively.}
%\end{minipage}
\label{Fig 2}
\end{figure*}
%%%%%%%%%%%%%%%%%%%%%%%%%%%%%%%%%%%%%%%%%%%%%%%%%%%%%%%%%%%%%%%%%%%%

When a nonpolarized electron flow passes through a conductor, a transverse pure spin current will be generated due to the spin orbit coupling (SOC). The spin current density can be read as
\begin{equation}
%\stackrel{\rightarrow}{J_{s}}=\theta_{SH}\frac{\hbar}{2e}\stackrel{\rightarrow}{J_{e}}\times\stackrel{\rightarrow}{\sigma},
\mathbf{J}_{s}=-\theta_{\text{SH}}\frac{\hbar}{2|e|}\mathbf{J}_{e}\times\boldsymbol{\sigma},
\label{h1vll}
\end{equation}
where $\theta_{\text{SH}}$ is the spin Hall angle, $\boldsymbol{\sigma}$ represents the spin, $\hbar$ is the reduced Planck constant, $e$ is the electronic charge, and $\mathbf{J}_{e}$ is the electron flow density~\cite{Hirsch1999,Takahashi2008}. In a cylindrical wire the spins wind around the surface. For a wire with rectangular cross-section like the commonest thin film sample, spins are accumulated at opposite interfaces with opposite spin directions, as shown in Fig. 1. On the discussion about YIG/Pt system up to now, spin accumulation at top and bottom interfaces, as shown in Fig. 1(a), is considered as the only reason to trigger the SMR. Hereinto, the magnetization direction of magnetic insulator layer (e.g. YIG) parallel with and perpendicular to accumulated spins at YIG/Pt interface in XY plane, $\mathbf{M}\parallel\boldsymbol{\sigma}$ and $\mathbf{M} \bot \boldsymbol{\sigma}$, are deemed to be two maximum states of spin scattering and spin absorption, respectively. The resistance of Pt film can therefore be tuned by $\mathbf{M}$ under the interaction of inverse SHE, causing the SMR effect. The influence of spin accumulation at left and right interfaces shown in Fig. 1(b) was neglected in the SMR. Being the most important conclusion from this physical image of the SMR, magnetoresistance must remain constant when $\mathbf{M}$ is rotating in XZ plane, because $\mathbf{M}$ has no transverse component. On the other hand, the magnetic insulator layer seems to be indispensable for the observation of SMR. This is the reason why so many studies on the SMR have chosen YIG, Fe$_3$O$_4$ and CoFe$_2$O$_4$ as the bottom layer~\cite{Hahn2013,Vlietstr2013APL,Marmion2014,Ding2014,Vliestra2013,Isasa2014}.

If spin splitting occurs in the Z direction, the spin accumulation described in Fig. 1(b) becomes a well-known spin transport phenomenon, anomalous Hall effect (AHE). Countless studies of the AHE have already proved the significant existence of the spin accumulation at the left and right interfaces. We hardly believe it could totally be missing in the SMR effect and thus we take it back in our model.
Fig. 2(a) exhibits the coordinate system for the SMR measurement. The yellow arrow ($\mathbf{J_e}$) indicates the electron flow direction. The red arrow ($\mathbf{M}$) indicates $\mathbf{H}$ direction and it is also the direction of saturation magnetic moment if $\mathbf{H}$ is large enough. The gray arrow is the Z direction. $\alpha$, $\beta$, and $\gamma$ represent rotation angles of $\mathbf{H}$ in different planes, respectively. When an electron flow is passing through a normal metal layer deposited on a magnetic layer, partial conduction electrons can be polarized with the spin direction along $\mathbf{M}$ direction by means of spin transfer torque, spin filtering and magnon-spin angle moment transfer etc.~\cite{STT,Wang2006,Lukashev2013,Zhang2012}. In this case $\mathbf{J_e}$ could be separated into polarized ($\mathbf{J_e^p}$) and nonpolarized ($\mathbf{J_e^{np}}$) parts. Still, spin accumulation will arise due to $\mathbf{J_e^{np}}$ as shown in Fig. 2(b). Meanwhile the polarized conduction electrons marked as $\mathbf{M}$ direction at the center of every image in Fig. 2(b), incline to move to a preferred direction in the light of SHE. If we consider the four spin accumulation interfaces as four preferred spin channels, the physical image indicated in Fig. 2(b) are then greatly similar to the typical dual-conduction-channel model of current in plane giant magnetoresistance (CIPGMR)~\cite{GMR}. Here, if the spin direction of an electron is parallel to the spin direction of one channel, it is easy for the electron to pass through this channel. On the contrary, antiparallel configuration means strong scattering. Perpendicular configuration means that the spin has projections in both parallel and antiparallel directions, which leads to middle-level electron scattering. Furthermore, for a film sample, usually its width (w) is much larger than its thickness (t), $\text{w}>>\text{t}$. Most conduction electrons with spin in Z direction can hardly approach the left and right channels. Oppositely, those conduction electrons with spin in Y direction can easily access the top and bottom channels since the spin diffusion length is usually larger than t. Therefore, Fig. 2(b) intuitively gives the typical relationship $\rho_{\parallel}\approx\rho_{\bot} >\rho_{T}$ in the SMR effect. Accordingly, Fig. 2(c) presents equivalent circuits for the above analysis.

For \textbf{M} in XY plane, the resistivity tensor can be written as (details see Appendix A)
\begin{equation}
\hat{\rho} =
\left(\begin{array}{cc}
\rho_{11} & \rho_{12}  \\
\rho_{21} & \rho_{22}
\end{array}
\right) =
\left(\begin{array}{cc}
\rho_{\parallel} & \epsilon   \\
\epsilon  & \rho_{T}
\end{array}
\right),
\end{equation}

Here $\rho_{11}$ and $\rho_{22}$ are longitudinal resistivities, $\rho_{12}$ and $\rho_{21}$ are transverse resistivities, respectively, when $\alpha$ = 0 and $90^\circ$. We make $\rho_{12}$ = $\rho_{21}$ = $\epsilon$ because they are usually minuteness. For arbitrary $\alpha$, using unitary transformation we get $\rho'=R_{-\alpha}\hat{\rho}R_{\alpha}$~\cite{Ranieri2008}, where
\begin{equation}
R_{\alpha} =
\left(\begin{array}{cc}
\cos{\alpha} & -\sin{\alpha}  \\
\sin{\alpha} &  \cos{\alpha}
\end{array}
\right),
\end{equation}
is the rotation matrix. Then the longitudinal resistivity in new coordinate system can be simply read as,
\begin{eqnarray}
\rho_{l}(\alpha) &=&
\left(\begin{array}{cc}
\cos{\alpha} & \sin{\alpha}  \\
\end{array}
\right)
\left(\begin{array}{cc}
\rho_{\parallel} & \epsilon \\
\epsilon  & \rho_{T}
\end{array}
\right)
\left(\begin{array}{c}
\cos{\alpha} \\
\sin{\alpha}
\end{array}
\right)  \notag\\
&=& \rho_{\parallel}\cos^2{\alpha}+2{\epsilon}\sin{\alpha}\cos{\alpha}+\rho_{T}\sin^2{\alpha}  \notag\\
&\approx&{\rho_{\parallel}}-(\rho_{\parallel}-\rho_{T})\sin^2{\alpha},
\label{vllpq}
\end{eqnarray}
Using our model, we get the same expression of the SMR on $\alpha$ as that in reference~\cite{Nakayama2013,Althammer2013,Chen2013}. Changing $\alpha$ to $\beta$ and $\gamma$, we can also obtain
\begin{eqnarray}
\rho_{l}(\beta)&\approx&\rho_{T}-(\rho_{T}-\rho_{\bot})\sin^2{\beta}, \notag\\
\rho_{l}(\gamma)&\approx&\rho_{\parallel}-(\rho_{\parallel}-\rho_{\bot})\sin^2{\gamma},
\end{eqnarray}
Therefore our expressions satisfy all rotation angles. This is distinctly different from the reported SMR mechanism.

Once the physical image of the SMR is clarified, we found that the magnetic insulator layer such as the YIG is not a requisite for the observation of SMR. The role of magnetic insulator layer is just for spin polarization of conduction electrons, and it actually works in a low efficiency manner. We therefore employ discontinuous magnetic metal layer to substitute for the magnetic insulator layer in this research.
%\section{Experimental details} %%%%%%%%%%%%%%%%%%%%%%%%%%%%%%%%%%%%%%%%%

Two samples, Co$_2$FeAl (0.3 nm)/Pt (2.5 nm) and Co$_2$FeAl (0.6 nm)/Pt (2.5 nm), were prepared with Hall bar mask on MgO substrates by DC sputtering at room temperature. The base pressure is $2 \times 10^{-5}$ Pa. The thicker sample was annealed at $320~^\circ$C \textit{in situ}. Magnetic transport properties of samples were measured by a physical property measurement system (PPMS). Ultrathin Co$_2$FeAl film cannot grow on MgO substrate uniformly, and thus it is insulating until the thickness is over 1.1 nm. Figure 1 shows schematic morphology of Co$_2$FeAl/Pt sample, where the isolated Co$_2$FeAl grains are enwrapped by the Pt layer. When conduction electrons pass through Co$_2$FeAl grains, partial electrons will be polarized along $\mathbf{M}$ direction because of angle moment exchanging. After that the polarized electrons move to the preferred channel under the influence of SHE.

%\section{Results and discussion} %%%%%%%%%%%%%%%%%%%%%%%%%%%%%%%%%%%%%%%%%

\begin{figure}
\centering
   \includegraphics[width=7 cm]{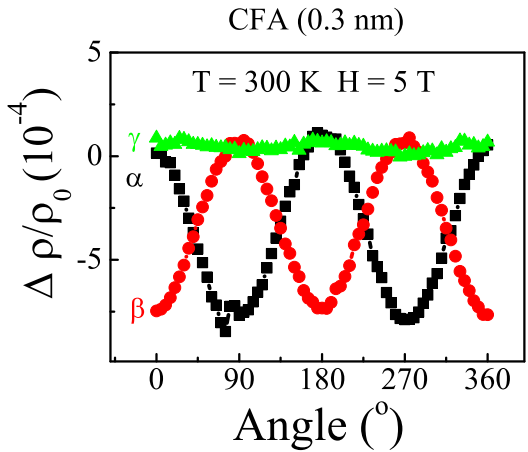}
   \caption{(Color online)  Field rotation magnetoresistance data with H = 5 T in three orthogonal planes for Co$_2$FeAl (0.3nm)/Pt (2.5 nm) at 300 K. $\rho_{0}$ is the resistivity at zero magnetic field.}
\label{Fig 3}
\end{figure}

Figure~\ref{Fig 3} shows classic SMR results of Co$_2$FeAl (0.3 nm)/Pt (2.5 nm) film under 5 T magnetic field at room temperature. $\Delta{\rho}$ = $\rho_{l} - \rho_{T}$ for $\alpha$. $\Delta{\rho}$ = $\rho_{l} - \rho_{\bot}$ for $\beta$ and $\gamma$. In this figure, the dependency of $\gamma$ is almost flat, which is thought as the most important evidence for the SMR based on the theory raised in Ref.~\cite{Chen2013}. However, it is still a function of $\cos^2{\gamma}$ as $\Delta{\rho}\approx(\rho_{\parallel}-\rho_{\bot})\cos^2{\gamma}$ according to our model. It looks flat just because $\rho_{\parallel}\approx{\rho_{\bot}}$, which is resulting from dimension effect as shown in Fig.~\ref{Fig 2}(b). The weak (cosine)$^2$ relationship is observed not only in our experiment, but also in Ref.~\cite{Lu2013PRB,Lin2014,Miao2014,Althammer2013,Chen2013}. Even in the first report of the SMR, the relationship had appeared in Fig. 4(e) if one reads the data carefully~\cite{Nakayama2013}. Moreover, since we use much more efficient method for spin polarization of conduction electrons and the thinner Pt layer, the SMR effect is drastically enhanced and it is at least one order larger than those reported previously. These results strongly support our understanding on the SMR. It is more important that larger effect may find it easier in possible applications.

\begin{figure}
\centering
   \includegraphics[width=7cm]{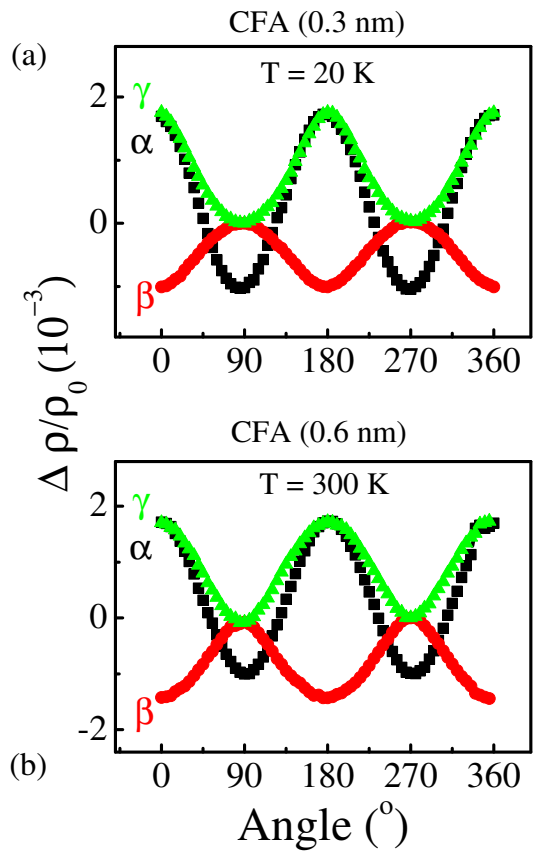}
   \caption{(Color online) Field rotation magnetoresistance data for (a) Co$_2$FeAl (0.3nm)/Pt (2.5 nm) at 20 K with H = 5 T and (b) Co$_2$FeAl (0.6nm)/Pt (2.5 nm) at 300 K, where H =1 T for $\alpha$ and H =2 T for $\beta$ and $\gamma$. $\rho_{0}$ is the resistivity at zero magnetic field.}
   \label{Fig 4}
\end{figure}

When $\rho_{\parallel}$ deviates from $\rho_{\bot}$, the $\cos^{2}\gamma$ dependence would become more distinct. There are two ways to enlarge the difference between $\rho_{\parallel}$ and $\rho_{\bot}$. Firstly, our micromechanism of the SMR is based on the SHE as usual. The SOC can be enhanced and the spin diffusion length can be elongated at low temperature, which leads to intensive spin accumulation and makes more polarized electrons with spin along Y and Z directions accessing the channels. However, the spin along X direction is always perpendicular to channels' preferred spin directions, the related electrons are still difficult to enter the channels and thus $\rho_{\parallel}$ may change little. With the same reason, increasing $\mathbf{J_e^{p}}$ may also reduce $\rho_{\bot}$, since more electrons with spin along Z direction can access left and right channels. On account of these deduction, the experimental results shown in Fig. 4 can be understood. In both subgraphs of Fig. \ref{Fig 4}, the (cosine)$^2$ relationships of $\gamma$ are clearly observed. Note that in some group, they consider (cosine)$^2$ relationships of $\gamma$ stem from anisotropy magnetoresistance (AMR)~\cite{Lu2013PRB,Lin2014}, because they adopted the former SMR model. We believe there is no AMR contribution at low temperature. However, with increasing thickness of Co$_2$FeAl layer, AMR will dominate magnetoresistance behavior finally.

The SMR effect is a fascinating conception. It is significant to study the related spin transport behavior and application in spintronics. However, the physical image of the SMR was not clear in previous studies. Here we suggest a model enlightened by the dual-conduction-channel mode of CIPGMR, where the SMR stems from the  scattering between polarized and nonpolarized conduction electrons under SHE, rather than the interaction between the accumulated spins and the magnetization of magnetic insulator layer. According to our model, $\rho_{\parallel}\approx\rho_{\bot}>\rho_{T}$ comes from dimension effect. Magnetic insulator layer, such as YIG, is actually a kind of bad buffer layer for the observation of SMR. More importantly, when $\mathbf{H}$ rotates in XZ plane, the SMR effect still follows (cosine)$^2$ relationship. Therefore, some reported results which are thought as the AMR are actually the SMR too. This research may help to build a precise theory on the SMR effect.

% Figure 3 %%%%%%%%%%%%%%%%%%%%%%%%%%%%%%%%%%%%%%%%%%%%%%%%%%%%%%%%%
%%%%%%%%%%%%%%%%%%%%%%%%%%%%%%%%%%%%%%%%%%%%%%%%%%%%%%%%%%%%%%%%%%%%

%Figure 4%%%%%%%%%%%%%%%%%%%%%%%%%%%%%%%%%%%%%%%%%%%%%%%%%%%%%%%%%%%%%%%%%%%%%%%%%%%%%%%%%

%\bigskip
\begin{acknowledgments}
We thank the beamline 08U1 at the Shanghai Synchrotron Radiation Facilities (SSRF) for the sample preparation and measurement. This work was supported by the National Science Foundation of China Grant Nos. 51331004, 51171148, 11374228, 11205235, the National Basic Research Program of China under grant No. 2015CB921501 and the Innovation Program of Shanghai Municipal Education Commission No. 14ZZ038. Z. G. Zhu is supported by Hundred Talents Program of The Chinese Academy of Sciences. 
\end{acknowledgments}

\newpage
\appendix
\begin{center}
\chapter{\textbf{Appendix A}}
\end{center}
If the direction of the magnetization $\mathbf{M}$ is coincide with the x direction, we have the relation $\mathbf{E}=\hat{\rho}\mathbf{j}$, i.e.
\begin{equation}
\left(
\begin{array}{c}
E_{x} \\
E_{y}
\end{array}\right)
=
\left(
\begin{array}{cc}
\rho_{xx}(\mathbf{M}\parallel\mathbf{x}) &  \rho_{xy}(\mathbf{M}\parallel\mathbf{x})\\
\rho_{yx}(\mathbf{M}\parallel\mathbf{x}) &  \rho_{yy}(\mathbf{M}\parallel\mathbf{x})
\end{array}\right)
\left(
\begin{array}{c}
j_{x} \\
j_{y}
\end{array}\right),
\label{exey}
\end{equation}
where $E_{x(y)}$ is the electric field in $x(y)$ direction, $\rho_{ij}(\mathbf{M}\parallel\mathbf{x})$ means the driven current is in the $j$ direction, and the response electric field is along $i$ direction. $\rho_{yy}(\mathbf{M}\parallel\mathbf{x})$ indicates the longitudinal measurement (both the driving current and the response electric field are in $y$ direction) with a transverse magnetization (in $x$ direction), which is equivalent to $\rho_{T}$ as the definition in this work. And $\rho_{xx}(\mathbf{M}\parallel\mathbf{x})=\rho_{\parallel}$. $\rho_{xy}(\mathbf{M}\parallel\mathbf{x})$ and $\rho_{yx}(\mathbf{M}\parallel\mathbf{x})$ are the transverse response to the longitudinal driving current, which are just planar Hall resistivities. We assume that they are small quantities, labeled by $\epsilon$. We thus obtain the resistivity matrix
\begin{equation}
\hat{\rho}=\left(
\begin{array}{cc}
\rho_{\parallel} & \epsilon \\
\epsilon & \rho_{T}
\end{array}
\right).
\label{rhomatrix}
\end{equation}
If the magnetization deviates an angle $\alpha$ in xy plane from the $x$ direction, we can select the direction of the magnetization as the new $\mathbf{x}$ direction, i.e. $\mathbf{x}'$, which means we rotate the old coordinate to the new coordinate. With this coordinate transformation, one vector in the new coordinate $\mathbf{A}'$ has a relationship with the one in the old coordinate $\mathbf{A}$,
\begin{equation}
\mathbf{A}'=R^{-1}(\alpha)\mathbf{A}=R(-\alpha)\mathbf{A}.
\label{aprime}
\end{equation}
We therefore have $R(\alpha)\mathbf{E}'=\mathbf{E}$ and $R(\alpha)\mathbf{j}'=\mathbf{j}$. Taking account into these relations, we obtain
\begin{equation}
\mathbf{E}=\hat{\rho}\mathbf{j}\Rightarrow R(\alpha)\mathbf{E}'=\hat{\rho}R(\alpha)\mathbf{j}'\Rightarrow \mathbf{E}'=R^{-1}\hat{\rho}R\mathbf{j}'.
\label{Etransformed}
\end{equation}
In the new coordinate, a driving current in $\mathbf{x}'$ direction may cause an electric field in the same direction, leading to a longitudinal resistivity from the relation
\begin{equation}
E'_{x}=(1,0)\left(
\begin{array}{c}
E'_{x} \\
E'_{y}
\end{array}
\right)=\underbrace{(1,0)R^{-1}(\alpha)\hat{\rho}R(\alpha)
\left(
\begin{array}{c}
1  \\
0
\end{array}
\right)}_{\rho_{l}}
j.
\label{exprime}
\end{equation}
From Eq. (\ref{exprime}), we get the longitudinal resistivity
\begin{equation}
\rho_{l}=(\cos\alpha,\sin\alpha)\hat{\rho}\left(
\begin{array}{c}
\cos\alpha \\
\sin\alpha
\end{array}
\right).
\end{equation}
This will lead us the Eq. (4) in the main text.
At the same time, we obtain
\begin{eqnarray}
E'_{y} &=& (0,1)\left(
\begin{array}{c}
E'_{x} \\
E'_{y}
\end{array}
\right)=
(0,1)R^{-1}(\alpha)\hat{\rho}R(\alpha)j
\left(
\begin{array}{c}
1  \\
0
\end{array}
\right)  \notag\\
&=&\underbrace{(-\sin\alpha,\cos\alpha)\hat{\rho}\left(
\begin{array}{c}
\cos\alpha \\
\sin\alpha
\end{array}
\right)}_{\rho_{\text{Trans}}}j.
\label{eyprime}
\end{eqnarray}
Explicitly, we write the transverse resistivity
\begin{equation}
\rho_{\text{Trans}}=\epsilon\cos2\alpha+(\rho_{T}-\rho_{\parallel})\sin\alpha\cos\alpha.
\label{rhotrans}
\end{equation}

\end{document}